%% file: zsl_l1_foundation.tex
\documentclass{SciPost}

\hypersetup{
    colorlinks,
    linkcolor={red!50!black},
    citecolor={blue!50!black},
    urlcolor={blue!80!black}
}

\usepackage[bitstream-charter]{mathdesign}
\DeclareSymbolFont{usualmathcal}{OMS}{cmsy}{m}{n}
\DeclareSymbolFontAlphabet{\mathcal}{usualmathcal}

\usepackage{booktabs}
\usepackage{multirow}
\usepackage{placeins}
\usepackage{subcaption}

\newcommand{\fr}{\mathrm{FR}}
\newcommand{\lone}{\mathrm{L1}}

\newcommand{\pt}{p_{\mathrm{T}}}
\newcommand{\collide}{\textsc{Collide-2V}}
\newcommand{\minbias}{minimum-bias}
\newcommand{\ebops}{\mathrm{EBOPs}}

\fancypagestyle{SPstyle}{
\fancyhf{}
\lhead{\colorbox{scipostblue}{\bf \color{white} ~SciPost Physics }}
\rhead{{\bf \color{scipostdeepblue} ~Submission }}

\fancyfoot[C]{\textbf{\thepage}}
}

\begin{document}

\pagestyle{SPstyle}

\begin{center}{\Large \textbf{\color{scipostdeepblue}{
Real-time event-level zero-shot learning at 150\,ns:\\
towards a foundation model for particle physics experiment triggering\\
}}}\end{center}

\begin{center}\textbf{
Leo Rozanov\textsuperscript{1$\star$},
Lauri Laatu\textsuperscript{1},
Benedikt Maier\textsuperscript{1},
Arianna Cox\textsuperscript{1},
Aadeesh Sharma\textsuperscript{1},
Zhiqiang Que\textsuperscript{2} and
Alex Tapper\textsuperscript{1}
}\end{center}

\begin{center}
{\bf 1} Imperial College London, London, United Kingdom
\\
{\bf 2} University of Bristol, Bristol, United Kingdom

\footnotesize $\star$ Corresponding Author
\end{center}

\section*{\color{scipostdeepblue}{Abstract}}
\textbf{\boldmath{%
The hardware-based Level-1 triggers of the Large Hadron Collider (LHC) experiments select collisions with algorithms designed for one signature each, acting on a coarse, low-latency view of every event. We present a learned event embedding for several trigger tasks and evaluate its transfer to processes absent from training. Using paired offline and Level-1 views of the same simulated collisions, we train an offline transformer with a supervised metric loss and distil it, event by event, into a quantised Level-1 transformer whose bit widths are learned under a fixed resource budget. Trained on 25 Standard Model processes, the embedding places 28 held-out processes next to their physical relatives. A nearest-neighbour selection against background collisions, which never enter encoder training, rejects 99\% of them. It retains a median of 90\% of seen and 96\% of unseen Standard Model events and 88\% of a set of 52 beyond-the-Standard-Model benchmark samples, the best-performing score across the three groups among the baselines evaluated on the same events. A second gate against the seen processes isolates high-mass dilepton signals into label-free clusters, and long-lived signals too when displaced-track information is available. With 7\,512 parameters and an FPGA latency of 147\,ns, the encoder uses one event representation, learned once, to derive several selections, opening the door to a foundation model for the LHC's hardware-based triggers.
}}

\vspace{\baselineskip}

\noindent\textcolor{white!90!black}{%
\fbox{\parbox{0.975\linewidth}{%
\textcolor{white!40!black}{\begin{tabular}{lr}%
  \begin{minipage}{0.6\textwidth}%
    {\small Copyright attribution to authors. \newline
    This work is a submission to SciPost Physics. \newline
    License information to appear upon publication. \newline
    Publication information to appear upon publication.}
  \end{minipage} & \begin{minipage}{0.4\textwidth}
    {\small Received Date \newline Accepted Date \newline Published Date}%
  \end{minipage}
\end{tabular}}
}}
}

\vspace{10pt}
\noindent\rule{\textwidth}{1pt}
\setcounter{tocdepth}{1}
\tableofcontents
\noindent\rule{\textwidth}{1pt}
\vspace{10pt}

\section{Introduction}
\label{sec:intro}

The ATLAS and CMS experiments at the Large Hadron Collider (LHC) select collisions in real time, with a first trigger level implemented in custom hardware and field-programmable gate arrays (FPGAs), followed by a software trigger~\cite{ATLAS:2017tdaq,CMS:2020cmk}. At the High-Luminosity LHC (HL-LHC) the Level-1 (L1) trigger of CMS will reduce the 40\,MHz bunch-crossing rate to 750\,kHz within a latency of 12.5\,$\mu$s~\cite{CMS:2020cmk}. Only a few microseconds of this budget are available to any single algorithm, and every event rejected at this stage is lost. For the HL-LHC, CMS brings tracker information into L1 and combines it with calorimeter and muon information in a correlator trigger, which runs a simplified particle-flow reconstruction with per-particle pileup weights (PUPPI)~\cite{CMS:2020cmk,Bertolini:2014bba}. L1 algorithms can then act on a list of reconstructed particles rather than on calorimeter towers and muon stubs alone. The particle list is, however, a degraded version of the offline reconstruction, meaning tracking is restricted in angle, resolutions are coarser, and thresholds are higher.

Trigger menus are collections of algorithms, each designed for one signature. Machine-learned triggers so far follow the same pattern. The anomaly-detection triggers AXOL1TL and CICADA of CMS~\cite{Govorkova:2021utb,Zipper:2023axo,Gandrakota:2024axo,CMS:2026ekg}, jet taggers~\cite{Laatu:2025tft} and $\tau$ identifiers each have their own model, training sample and share of the FPGA. An alternative is a single encoder that maps every event to one vector from which several decisions are derived by cheap downstream operations. Offline, this is the premise of foundation models for particle physics~\cite{Harris:2024sra,Birk:2024knn,Golling:2024abg,Mikuni:2024qsr} where a representation is pre-trained once and reused for many tasks. We apply the same principle to create an event-level encoder that is small enough for the trigger and reused without retraining.

We use the encoder beyond the data it was trained on with zero-shot learning~\cite{Xian:2018zsl}. A model is trained on a set of \emph{seen} classes and evaluated on \emph{unseen} ones. We use the term in its class-holdout sense, asking whether the representation learned from seen processes is still useful for unseen ones, and whether the same encoding can support multiple trigger tasks, including background rejection and anomaly detection for both Standard Model (SM) and beyond-the-Standard-Model (BSM) processes.

The obstacle to learning such an embedding at L1 is the coarse view of L1 data relative to the offline view. The \collide{} dataset~\cite{Moreno:2025collide} provides both the offline and the L1 view of the same simulated collisions. Using this unique dataset, we're able to apply knowledge distillation~\cite{Hinton:2015kd} through a contrastive objective~\cite{Tian:2020crd} to supervise the L1 view with the offline view event by event. The closest precedent in particle physics, RS3L~\cite{Harris:2024sra}, pairs two re-simulations of the same jet; here the two views are two detector reconstructions of one collision, encoded by networks of different capacity.

Transformers for jet tagging have recently been implemented on FPGAs at latencies of order 100\,ns (Ref.~\cite{Laatu:2025tft} and references therein), using hls4ml~\cite{Duarte:2018ite,Schulte:2025mai} with high-granularity quantisation (HGQ)~\cite{Sun:2024hgq}, which learns a bit width for every weight and activation, together with distributed-arithmetic optimisation~\cite{Sun:2025da4ml}. We use HGQ and distributed arithmetic, through the Alkaid compiler~\cite{Sun:2026alkaid}, and move from jet-level tagging to an event-level embedding.

\begin{figure}[tbp]
  \centering
  \begin{subfigure}{\linewidth}
    \centering
    \includegraphics[width=\linewidth]{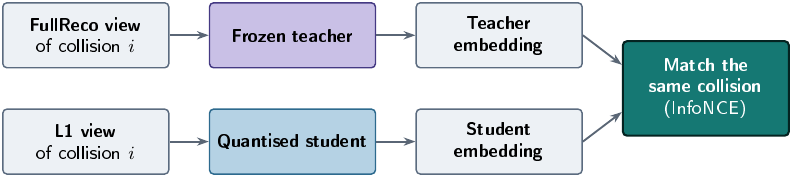}
    \caption{A frozen FullReco teacher supplies paired-event targets to the trainable HGQ L1 student.}
    \label{fig:overview-training}
  \end{subfigure}

  \medskip
  \begin{subfigure}{\linewidth}
    \centering
    \includegraphics[width=\linewidth]{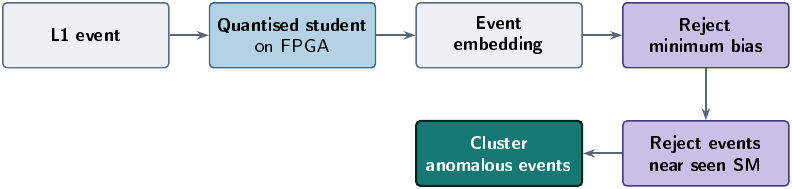}
    \caption{The event embedding supports \minbias{} rejection, seen-SM novelty selection and population clustering.}
    \label{fig:overview-analysis}
  \end{subfigure}
  \caption{Paired-view training (top) and use (bottom) of the event embedding. In Stage~B a symmetric InfoNCE loss, with an HGQ resource penalty, matches the student's embedding to the frozen teacher's embedding for each collision. The gates and clustering of Sec.~\ref{sec:trigger} are evaluated offline on the embedding.}
  \label{fig:overview}
\end{figure}

We train an offline transformer with a supervised triplet loss and distil it, event by event, into a quantised L1 transformer whose bit widths are learned under a fixed resource budget (Fig.~\ref{fig:overview}). With this embedding and no retraining, we then reject \minbias{} collisions, the generic soft interactions that make up almost all of the input rate, and apply a two-stage selection of events unlike any seen SM process to find and flag BSM processes. We also give an FPGA implementation of the encoder to perform under trigger conditions.

\section{Dataset and experimental setup}
\label{sec:setup}

\subsection{The \collide{} dataset}
\label{sec:dataset}
\collide{}~\cite{Moreno:2025collide} is a public sample of about 750 million simulated proton--proton collision events, generated with \textsc{MadGraph}~\cite{Alwall:2014hca} and \textsc{Pythia}~\cite{Bierlich:2022pfr} and passed through \textsc{Delphes}~\cite{deFavereau:2013fsa} under HL-LHC conditions with an average of 200 simultaneous proton--proton interactions per bunch crossing (pileup). Each event is stored in two views. The \emph{offline} view (FullReco, FR) is a CMS-like full reconstruction obtained with a tuned \textsc{Delphes} card. The \emph{L1} view is produced by a separate parameterisation of the L1 trigger reconstruction with degraded vertex, track and calorimeter performance, tracking restricted in pseudorapidity, higher $\pt$ thresholds and a correspondingly modified pileup weighting. In both views we use the list of PUPPI~\cite{Bertolini:2014bba} particle candidates, stored in descending $\pt$.

We use the 53 SM processes of the catalogue and 52 beyond-the-Standard-Model (BSM) benchmark samples. The SM set spans Higgs, di-Higgs, vector-boson, top-quark and QCD production, $\Upsilon\to\ell\ell$ and \minbias{}. The BSM benchmarks include hidden-valley dileptons, prompt and displaced $R$-parity-violating supersymmetry, SUEP-like showers, $H\to aa$ decays into $b$ quarks, $\tau$ leptons or photons, a 500\,GeV $Z^{\prime}\to q\bar{q}$, and seven light-scalar samples of Ref.~\cite{DAgnolo:2025cxb}, one of which is a control sample. Several samples have displaced decays with proper decay lengths of 1 to 100\,mm. The full lists with event counts are given in Appendix~\ref{app:classes}. The SM samples used here contain between 200\,000 and 1.5 million events each (median 1.41 million, 70.2 million in total); the BSM samples contain about 50\,000 events each. Classes are sampled with equal weight rather than by cross-section, so all acceptances and retentions refer to balanced process mixtures, not to a physical trigger rate (Sec.~\ref{sec:res-minbias}).

\subsection{Inputs and preprocessing}
\label{sec:preproc}
Each event is represented by its PUPPI candidates with 14 features per particle: transverse momentum $\pt$, pseudorapidity $\eta$, azimuthal angle $\phi$, charge, mass, the transverse and longitudinal impact parameters $d_0$ and $d_z$ and their uncertainties, and one-hot particle-type flags for charged hadrons, neutral hadrons, photons, electrons and muons. The impact parameter $d_0$ is that of the true particle trajectory, without resolution effects or an acceptance limit, so that the L1 view keeps charged candidates with $|d_0|$ of several centimetres in the long-lived samples. The baseline track reconstruction primarily reconstructs prompt tracks; displaced tracks require an extended mode with reduced efficiency~\cite{CMS:2020cmk,Ryd:2020ear}. The longitudinal parameter $d_z$ is measured from $z=0$ and mostly reflects the position of the primary vertex. Candidates with zero PUPPI weight (pileup candidates, more than 95\% of the stored L1 list) and candidates with $\pt\le2$\,GeV are removed. The teacher reads the 256 leading FR particles and the student the 32 leading L1 particles, with shorter lists zero-padded.

The FR and L1 views of the same collision form a pair, and pairs in which either view is empty after the selection are dropped. About a third of the \minbias{} events, 37\% of the Drell--Yan events and a fifth of the $\Upsilon$ events have no L1 candidate that survives the selection, so these processes are represented by their more active events only.

\subsection{Seen/unseen split and samples}
\label{sec:split}
The SM processes are split at random into 25 seen and 28 unseen classes (Table~\ref{tab:classes-sm}). The protocol requires \minbias{} to be unseen, so that the gate of Sec.~\ref{sec:trigger} is calibrated on a process the encoders have never seen. This split satisfies that requirement. Both training stages use only the seen classes, so \minbias{} events enter only as reference and calibration samples at evaluation time (Sec.~\ref{sec:trigger}). All 52 BSM samples are used only for evaluation.

Each file of the dataset is split by row into a training part (the first 90\%) and a validation part (the last 10\%). Teacher and student are trained on the same class-balanced subset of the paired training rows, with 168\,990 event pairs per seen class (4.22 million in total) and a validation subset of 18\,340 pairs per class. All models are evaluated on the same 1\,000 events per SM and BSM process and 10\,000 \minbias{} events.

\section{Method}
\label{sec:method}

\subsection{Encoder architectures}
\label{sec:arch}
\begin{figure}[tbp]
  \centering
  \includegraphics[width=0.95\linewidth,height=0.74\textheight,keepaspectratio]{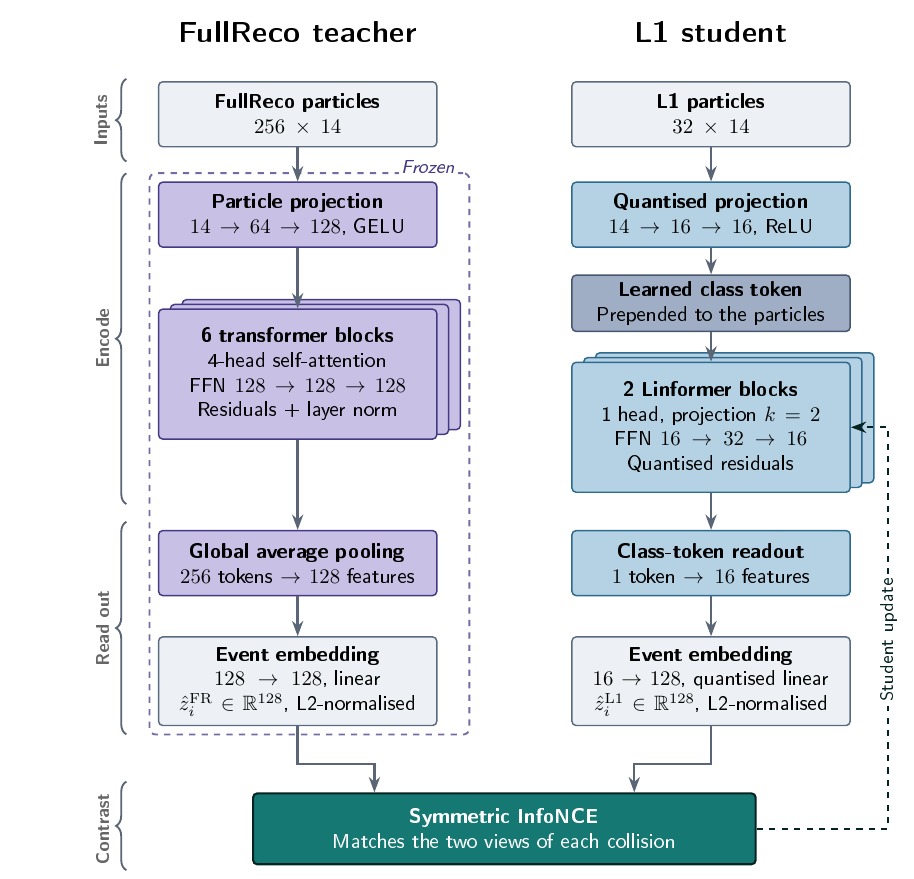}
  \caption{Encoder architectures and the distillation loss. The FullReco teacher (dashed box) is trained first with a triplet loss and then frozen. The symmetric InfoNCE loss pulls the student's embedding of each L1 event towards the teacher's embedding of the same collision and away from the other events in the batch. Only the student is updated, with an HGQ resource penalty.}
  \label{fig:training}
\end{figure}

We train two transformer encoders~\cite{Vaswani:2017attn}, a teacher and a student, each acting on the particle set (Fig.~\ref{fig:training}, Table~\ref{tab:arch}). Per-particle features are projected to the model width by two dense layers, passed through attention and feed-forward blocks with residual connections, and reduced to a single event vector. The student's Linformer projections act on the particle slots of the $\pt$-ordered list, as in Ref.~\cite{Laatu:2025tft}.

The \emph{teacher} is a floating-point model with feed-forward blocks of width 128 and GELU activations. Its event vector is the global average over the 256 particle slots, followed by a linear projection to 128 dimensions. The \emph{student} has width 16, two blocks, single-head Linformer attention~\cite{Wang:2020linformer} projecting keys and values to $k=2$ sequence rows, a $16\to32\to16$ feed-forward network, ReLU activations and no normalisation. A learned class token~\cite{Devlin:2019bert,Dosovitskiy:2021vit} is quantised and prepended to the 32 particle tokens, so the attention stack acts on 33 tokens. The event vector is the output of this token after the last block, projected linearly to 128 dimensions. The class-token readout is used instead of global average pooling, which in the fixed-length, zero-padded input would average over padding slots and require a division in fixed point. Because the final projection is linear from a 16-dimensional token, the student's 128-dimensional output lies in a 16-dimensional affine subspace. All 7\,512 student parameters and the activations are quantised with HGQ (Sec.~\ref{sec:hgq}). The trained student reaches $4.93\times10^{5}$ effective bit operations (EBOPs), with a mean bit width of 2.4 bits over its 29\,376 learned quantisers and a median quantiser width of 1.1 bits. Both event vectors are normalised to unit length (L2-normalised) before every loss and every evaluation, so all distances are chord distances on the unit sphere and all similarities are cosines.

\begin{table}[t]
  \centering
  \begin{tabular}{lcc}
    \toprule
     & Teacher & Student \\
    \midrule
    Input view            & FullReco & L1 \\
    Input particles       & 256      & 32 \\
    Model width           & 128      & 16 \\
    Transformer blocks    & 6        & 2 \\
    Attention             & multi-head, 4 heads & Linformer, 1 head, $k=2$ \\
    Normalisation         & layer norm (post-residual) & none \\
    Readout               & global average pooling & learned class token \\
    Arithmetic            & floating point & HGQ fixed point \\
    Parameters    & 623\,296 & 7\,512 \\
    EBOPs (trained)       & --       & $4.93\times10^{5}$ \\
    \bottomrule
  \end{tabular}
  \caption{Teacher and student configurations. Bit widths are learned per parameter and activation (Sec.~\ref{sec:hgq}).}
  \label{tab:arch}
\end{table}

\subsection{Stage A: supervised metric learning of the teacher}
\label{sec:stageA}
The teacher $f_\theta$ is trained on the offline view of the seen classes with a triplet loss~\cite{Schroff:2015facenet}. For an anchor $a$, a positive $p$ from the same class and a negative $n$ from a different class,
\begin{equation}
  \mathcal{L}_{\mathrm{A}} = \max\!\big(0,\; \lVert \hat{z}_a - \hat{z}_p \rVert_2 - \lVert \hat{z}_a - \hat{z}_n \rVert_2 + \alpha \big),
  \qquad \hat{z} = \frac{f_\theta(x^{\fr})}{\lVert f_\theta(x^{\fr}) \rVert_2},
  \label{eq:triplet}
\end{equation}
with margin $\alpha = 0.5$. Triplets are formed on the fly from a class-stratified stream. The anchor and positive are two events from a per-class reservoir, and the negative is drawn from another class with probability proportional to its number of buffered events, with no hard-negative mining. The teacher is trained for 50 epochs with batches of 128 triplets and AdamW~\cite{Loshchilov:2019adamw} at a learning rate of $10^{-5}$ (Appendix~\ref{app:training}).

We train the teacher with a triplet loss because a metric loss acts directly on distances in the embedding, which is the geometry the student inherits. In a preliminary comparison on five classes, it gave tighter and more ordered class regions than a cross-entropy classifier. We did not optimise this choice further, as the focus here is the distillation and its FPGA deployment.

\subsection{Stage B: event-paired distillation to Level-1}
\label{sec:stageB}
With the teacher frozen, the student $g_\phi$ embeds each L1 event to match its paired teacher embedding and differ from the others in the batch. With $s_{ij} = \hat{z}^{\fr}_i \cdot \hat{z}^{\lone}_j / \tau$, the symmetric InfoNCE loss~\cite{Oord:2018cpc} is
\begin{equation}
  \mathcal{L}_{\mathrm{B}} = -\frac{1}{2N}\sum_{i=1}^{N}
  \left[ \log \frac{e^{s_{ii}}}{\sum_{j=1}^{N} e^{s_{ij}}}
       + \log \frac{e^{s_{ii}}}{\sum_{j=1}^{N} e^{s_{ji}}} \right],
  \label{eq:infonce}
\end{equation}
with temperature $\tau = 0.05$ and batch size $N=128$. No class labels enter Stage~B. The negatives are the other events in the batch regardless of their class. With 25 balanced classes, about 4\% of the negatives share the anchor's class. The loss therefore asks the student to resolve individual events, not only classes. Appendix~\ref{app:alignment} shows that it does so for seen and unseen processes alike. Each student is trained for up to 250 epochs with AdamW at a learning rate of $10^{-5}$.

\subsection{Quantisation-aware training under a resource budget}
\label{sec:hgq}
The student is trained with HGQ~\cite{Sun:2024hgq}, which attaches a fixed-point quantiser to every weight and activation and learns the bit widths by gradient descent together with the weights. The training objective is
\begin{equation}
  \mathcal{L} = \mathcal{L}_{\mathrm{B}} + \beta \cdot \overline{\ebops},
  \label{eq:hgq}
\end{equation}
where $\overline{\ebops}$ is a differentiable estimate of the model's EBOPs, a proxy for FPGA logic usage that counts each multiplication with the product of its operand bit widths. The coefficient $\beta$ is adjusted once per epoch by a proportional--integral controller that drives the model towards a target of $5\times10^{5}$ EBOPs, raising $\beta$ while the model is above the target and halving it once the target is reached (Appendix~\ref{app:training}). The reported checkpoint is the epoch with the lowest validation loss among the epochs that satisfy the budget. Appendix~\ref{app:teacher-ablation} compares two attention-quantiser configurations. The headline student uses the first (Appendix~\ref{app:training}).

\subsection{Trigger functions on the embedding}
\label{sec:trigger}
All decisions use the L2-normalised student embedding and labelled reference sets separate from the evaluated events. They are computed offline on cached embeddings. 

\paragraph{Minimum-bias gate.}
An event is scored by the mean Euclidean distance to its $k=20$ nearest neighbours in a reference set of \minbias{} embeddings, a nearest-neighbour outlier score~\cite{Ramaswamy:2000knn,Sun:2022knn}, where a larger score means less \minbias{}-like. The threshold is the $(1-r)$ quantile of the scores of a separate calibration set of \minbias{} events, so that a fraction $r$ of them pass. Since \minbias{} is an unseen class, no \minbias{} event was used by the encoders or by the normalisation statistics. The gate uses 653\,661 \minbias{} events from files never used in training, split ten times at random into 50\,000 reference, 50\,000 calibration and 553\,661 test events, and we quote the acceptance measured on the test events. Signal retention is measured on 500 events per process. The choice of $k$ is examined in Appendix~\ref{app:diagnostics}.

\paragraph{Seen-SM novelty gate.}
Events that pass the \minbias{} gate are scored against each of the 25 seen classes. For seen class $c$ with a reference set $R_c$ and a calibration set $C_c$, let $d_c(x)$ be the distance from $x$ to its $k$-th nearest neighbour in $R_c$, with $k=10$. The score is standardised per class on the logarithmic scale,
\begin{equation}
  s_c(x) = \frac{\log d_c(x) - \mathrm{median}_{y\in C_c}\,\log d_c(y)}{\mathrm{IQR}_{y\in C_c}\,\log d_c(y)/1.349},
  \qquad s(x) = \min_c s_c(x),
  \label{eq:novelty}
\end{equation}
so that broad and narrow seen classes are comparable. The score $s(x)$ measures how far $x$ is from the \emph{closest} seen class in units of that class's spread. The threshold on $s$ is set so that 1\% of the \minbias{} calibration events pass it, separately from the threshold of the first gate. An event passes both gates if it is far from \minbias{} and from every seen class. The selection is evaluated on 300 events per process, with reference and calibration sets of 400 and 200 other events per seen class and 3\,000 each of \minbias{}.

\paragraph{Clustering.}
The survivors are projected with a principal-component analysis~\cite{Pedregosa:2011sklearn} fitted on the seen reference embeddings (32 components) and clustered with HDBSCAN~\cite{McInnes:2017hdbscan} (minimum cluster size 20, minimum samples 10). No labels enter either step. These settings were fixed before the analysis, and Appendix~\ref{app:discovery} studies their effect. Labels are used only to score the result, through the process composition of each cluster and a per-sample score defined in Appendix~\ref{app:discovery}.

\subsection{Baselines}
\label{sec:baselines}
We compare five scores on the same L1 events at matched \minbias{} acceptance: $H_{\mathrm{T}}$ (the scalar $\pt$ sum of the candidates), candidate multiplicity, a nearest-neighbour score on 18 hand-built event features, the same score on the student's $32\times14$ inputs, and an autoencoder reconstruction error. The nearest-neighbour scores use $k=20$. $H_{\mathrm{T}}$, multiplicity and the hand-built-feature score use all selected candidates, while the input-level score and autoencoder use the student's first 32 particles. Implementation details are given in Appendix~\ref{app:diagnostics}.

The autoencoder is trained only on \minbias{} events. Seen SM, unseen SM and BSM events are evaluated without retraining, with seen and unseen defined by the embedding's training split. Each score is thresholded on an independent \minbias{} calibration sample. The offline FullReco teacher is shown separately as a reference. We also train the same HGQ student without a teacher, using the Stage~A triplet loss on the seen L1 classes with the same data, resource target and initialisation seed.

\FloatBarrier

\section{Results}
\label{sec:results}

\subsection{Minimum-bias rejection}
\label{sec:res-minbias}
At 1\% \minbias{} acceptance, that is 99\% rejection, the embedding retains a median of 90\% of the seen SM processes, 96\% of the unseen SM processes excluding \minbias{}, and 88\% of the BSM processes (Table~\ref{tab:baselines}), with a measured test acceptance of 0.99\%. Although the embedding is trained on 25 SM processes and its threshold is set on \minbias{} events it never saw, it retains the unseen SM and BSM processes as well as its training processes. The slightly higher unseen retention reflects the composition of the groups, which $H_{\mathrm{T}}$ shows too. Since all processes are sampled with equal weight, these are per-process efficiencies, not trigger rates.

\begin{figure}[tbp]
  \centering
  \includegraphics[width=\linewidth,height=0.82\textheight,keepaspectratio]{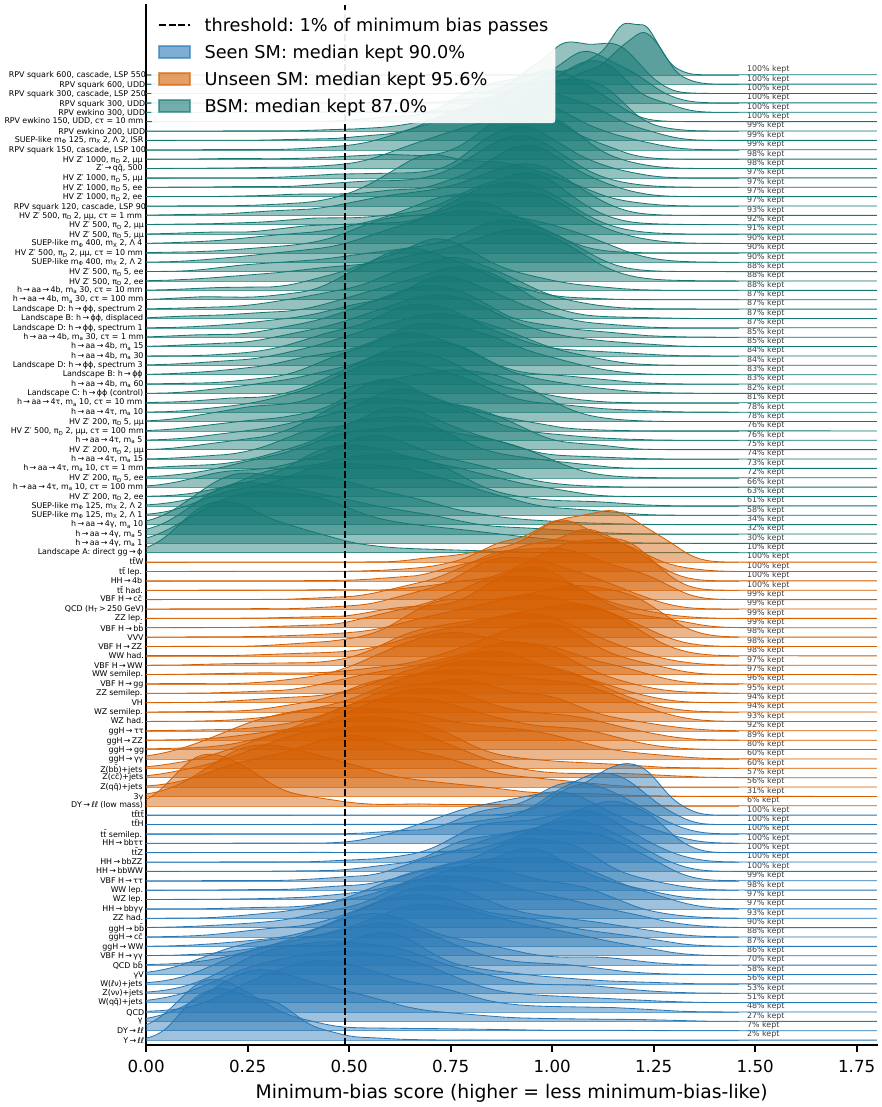}
  \caption{Distributions of the \minbias{} score, the mean distance to the 20 nearest \minbias{} reference events, for every seen SM (bottom), unseen SM (middle) and BSM (top) process, for one split, with 500 events per process. Within each group, processes are ordered by the fraction of events above the dashed threshold, which passes 1\% of the \minbias{} calibration events. The legend gives group medians for this split. Section~\ref{sec:res-minbias} gives averages over splits and Fig.~\ref{fig:minbias-processes} gives per-process retention.}
  \label{fig:minbias-scores}
\end{figure}

Most processes sit far from \minbias{} in the embedding (Fig.~\ref{fig:minbias-scores}). The tenth-percentile process retains 36\%, 57\% and 60\% of its events in the three groups, and the 90th percentile more than 99\%. The processes retained least have low particle multiplicity and final states dominated by photons or neutral particles, for which the L1 view carries no track information, such as $\Upsilon\to\ell\ell$ (2\%), tri-photon production (32\%) and $H\to aa\to4\gamma$ (30--35\%) (Fig.~\ref{fig:minbias-processes}).

\begin{figure}[tbp]
  \centering
  \includegraphics[width=\linewidth,clip]{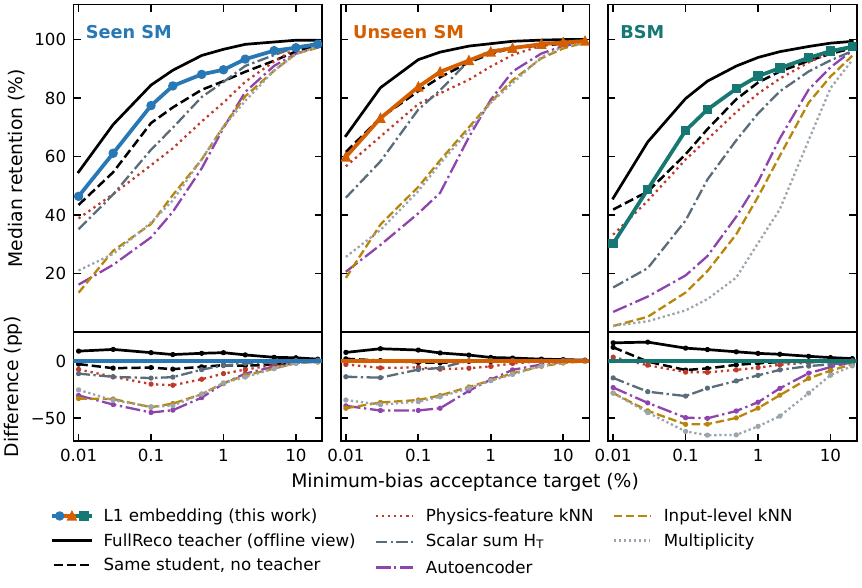}
  \caption{Comparison of the L1 embedding, the offline FullReco teacher, the same student trained without a teacher, and five baselines defined in Sec.~\ref{sec:baselines}. Columns show seen SM, unseen SM and BSM processes from left to right. The main panels give the median retention, averaged over the ten calibration splits of Sec.~\ref{sec:trigger}. The bottom panels give each comparison model's retention minus the L1-embedding retention at the same acceptance target. Positive values favour the comparison model.}
  \label{fig:baselines}
\end{figure}

The embedding is the best-performing score across the three groups (Table~\ref{tab:baselines}, Fig.~\ref{fig:baselines}). It exceeds $H_{\mathrm{T}}$ by 4 points on seen SM and 13 points on BSM processes and matches it on unseen SM processes, and it exceeds the hand-built features by 5 to 11 points on SM and 6 points on BSM processes. It retains far more than the autoencoder, the multiplicity and the same distance computed on the student's inputs, so the separation comes from the encoder rather than from the distance score. The FullReco teacher, which retains 94 to 99\% of every group on the offline view (Fig.~\ref{fig:baselines}), shows what the L1 view and the compression to 7\,512 parameters cost.

\input{tables_v3_single/baselines_table.tex}

\subsection{Selection of events unlike any seen process}
\label{sec:res-discovery}
The second trigger function selects events that are unlike \minbias{} and unlike every seen SM process. Both gates together pass 0.23\% of \minbias{} events and the novelty gate removes almost all SM events that pass the first gate. No seen process keeps more than 2.7\% of its events (median 1.0\%), and the unseen SM processes are removed almost as effectively (median 1.3\%, maximum 3.3\%), so the gate rejects SM processes that resemble seen ones and not only the 25 it was calibrated on.

\begin{figure}[tbp]
  \centering
  \includegraphics[width=\linewidth]{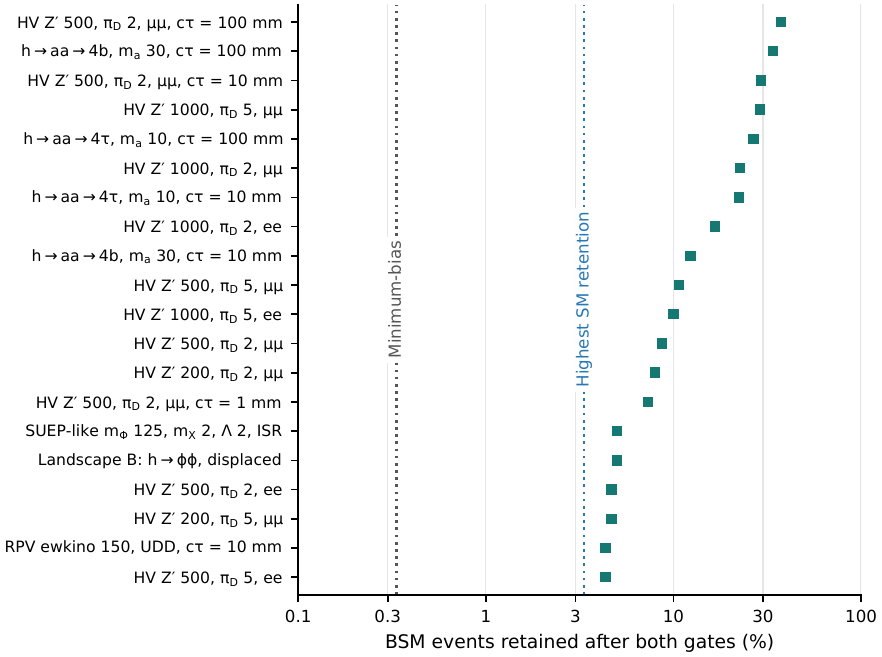}
  \caption{The 20 BSM processes with the highest fraction of events retained after both gates, out of 300 stream events per process. The blue dotted line marks the highest retention among seen and unseen SM processes; the grey dotted line marks the fraction of \minbias{} events passing both gates.}
  \label{fig:discovery-repeats}
\end{figure}

The 20 processes retained most are all BSM samples (Fig.~\ref{fig:discovery-repeats}) and fall into two groups. The long-lived hidden-valley and $H\to aa$ samples, with proper decay lengths of 10 or 100\,mm, are retained at 12 to 37\%, four to eleven times the highest SM retention. This relies on the impact parameters. With $d_0$ set to zero at inference, the $c\tau=100$\,mm samples fall to the SM level. The prompt high-mass dilepton samples, hidden-valley $Z^{\prime}$ models at 500 and 1\,000\,GeV, reach up to 29\%. The hadronic BSM samples pass the first gate but resemble the seen top-associated, di-Higgs and multijet processes, so the second gate removes them by design. A supervised selection on the same embedding would serve them.

The surviving events group by signature without labels. HDBSCAN finds a cluster of 327 events, 97\% of them from the prompt high-mass dilepton samples, at all 27 clustering settings tested. The long-lived events form a second cluster of 543 events, 80\% long-lived and 7\% SM (Appendix~\ref{app:discovery}). Samples that share a signature look alike at L1, so the clusters separate signatures rather than individual generator samples.

Both populations depend on distillation from the offline view (Table~\ref{tab:teacher}, Appendix~\ref{sec:res-ablation}). The same student distilled from an L1-view teacher keeps 8\% of the long-lived and 0.3\% of the high-mass dimuon events, against 24\% and 26\%, and forms no clusters. Without a teacher, more SM events pass the novelty gate (up to 6.3\% of a process) and the two populations do not form separate clusters. Which signatures survive also depends on the seen processes. With a second random split the long-lived samples are retained more and most prompt dimuon events are removed (Appendix~\ref{sec:res-split48}).

\subsection{Structure of the embedding}
\label{sec:res-structure}
\begin{figure}[tbp]
  \centering
  \includegraphics[width=\linewidth]{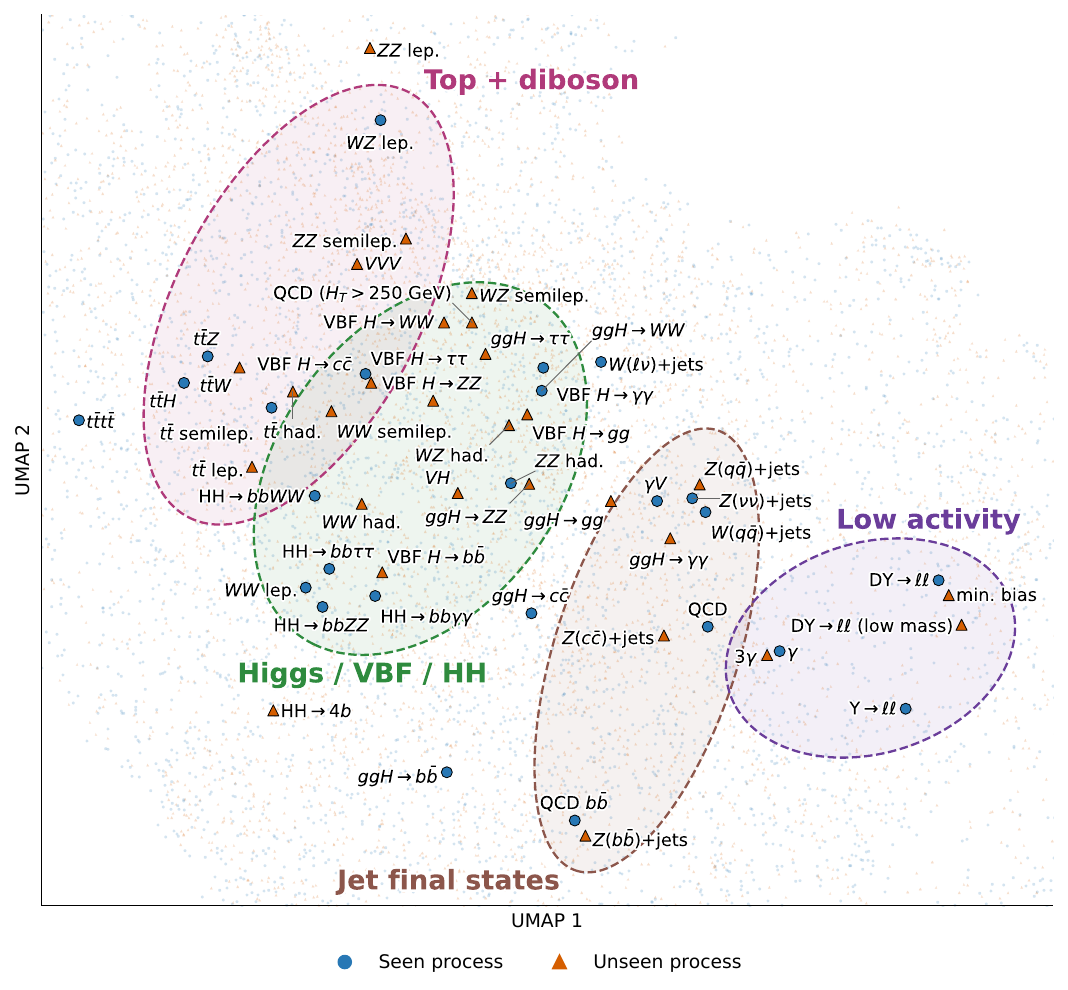}
  \caption{UMAP visualisation of the student embedding, with each event averaged with its mirror image along the beam axis to remove its orientation (Appendix~\ref{app:orientation}). Faint points show events, and labelled circles and triangles mark the medians of seen and unseen SM processes.}
  \label{fig:umap}
\end{figure}

Figure~\ref{fig:umap} gives a UMAP~\cite{McInnes:2018umap} visualisation of the student embedding for seen and unseen SM processes. Top-associated production and the leptonic dibosons lie on one side of the map, Higgs production through gluon fusion, vector-boson fusion and di-Higgs production in the centre, and the $W/Z$+jets and QCD processes next to them. The \minbias{} sample sits apart in a corner, together with the other low-activity processes: Drell--Yan, $\Upsilon\to\ell\ell$, and the photon samples. The 128-dimensional embedding shows the same pattern. The nearest centroid to unseen $ggH\to ZZ$ is the seen $ggH\to WW$, to $t\bar{t}W$ the seen $t\bar{t}Z$, and to unseen \minbias{} the seen Drell--Yan sample.

Unseen classes retain 89\% of the local separability of seen classes. A labelled-reference nearest-neighbour probe ($k=10$, 500 reference and 500 query events per class, applied within each group) reaches 20.6\% for the 25 seen and 18.2\% for the 28 unseen classes, against chance levels of 4.0\% and 3.6\% (Fig.~\ref{fig:separability}), about three quarters of the FullReco teacher's 28.7\% and 24.6\%. The absolute accuracies are modest because many seen processes differ only in the decay of one boson and are not separable event by event. The embedding nevertheless groups events by their physics content, and places processes it never saw almost as well as those it was trained on.

\section{FPGA implementation}
\label{sec:fpga}
The quantised student is converted to firmware with Alkaid~\cite{Sun:2026alkaid}, using the Linformer and multi-head attention support introduced in Ref.~\cite{Laatu:2025tft}. Constant matrix--vector products are implemented by distributed arithmetic~\cite{Sun:2025da4ml}. The target is an AMD Virtex UltraScale+ VU13P, from the FPGA family foreseen for the CMS Phase-2 correlator trigger~\cite{CMS:2020cmk}.

Table~\ref{tab:resources} gives the resource usage after placement and routing with Vivado 2025.2.1, for the \texttt{xcvu13p-flga2577-2-e} device. The design is fully unrolled, pipelined and placed entirely in one of the four super logic regions (SLRs), the separate silicon dies of the device. 
Its latency is 147\,ns. 
With a mean bit width below three bits, the arithmetic is implemented almost entirely in look-up tables. The distance scores of Sec.~\ref{sec:trigger} are evaluated offline, and distances to a few centroids or a learned head could provide compact firmware gates.

\input{tables_v3_single/resource_comparison_table.tex}

\section{Conclusion}
\label{sec:conclusion}
We have trained an event-level embedding for the Level-1 trigger by distilling an offline transformer into a quantised L1 transformer, using paired views of the same simulated collisions so that the offline reconstruction supervises the L1 reconstruction event by event. The embedding places unseen SM processes next to their physical relatives and retains 89\% of the local separability of the seen ones. A single nearest-neighbour score on it, calibrated on \minbias{} collisions the encoder has never seen, rejects 99\% of them. It retains a median of 90\% of seen SM, 96\% of unseen SM and 88\% of BSM processes, the best-performing score across the three groups compared with an autoencoder, an input-level score and kinematic baselines on the same events.

A second gate against the seen processes removes almost all SM events while retaining a large fraction of prompt high-mass dilepton signals. With displaced-track impact parameters at L1, it also retains a tenth or more of several long-lived signals. The survivors separate without labels into a 97\% pure prompt-dilepton population and a long-lived population of 80\% purity. Both populations rely on distillation from the offline view. With an L1-view teacher few high-mass dimuon events survive, and without a teacher the populations do not form separate clusters. The encoder fits in one SLR of a single FPGA of the kind used in the L1 trigger, with a latency of 147\,ns.

\paragraph{Outlook.}
The encoder is trained on a fraction of the available data and evaluated on two splits of the SM processes. Training on more data and processes, and on further splits, would show how far the transfer to unseen processes can be pushed. Neither the teacher's triplet loss nor the student's contrastive loss was fully optimised, and other objectives, larger teachers and other clustering algorithms could give a cleaner embedding and purer populations. Implementing the gates in firmware would allow the full chain from particles to decision to be measured, and the method must finally be tested on collision data, with zero-bias events, recorded without any trigger selection, as the reference. The teacher is trained on simulated process labels, and weak supervision on mixed samples of collision data~\cite{Metodiev:2017vrx} could reduce this dependence on simulation.

\section*{Acknowledgements}
We thank the authors of \collide{} for making the dataset available, Oz Amram for the BSM samples in \collide{}, and the developers of HGQ, hls4ml, da4ml and Alkaid for their tools and advice. We acknowledge computational resources and support provided by the Imperial College Research Computing Service (\url{http://doi.org/10.14469/hpc/2232}).

\paragraph{Funding information}
LL is funded by the Science and Technology Facilities Council (STFC) grant ST/W000636/1. BM acknowledges the support received through the STFC Ernest Rutherford Programme. LR, AC, and LL acknowledge the support of the COST Action EPIGRAPHY (CA24153).

\paragraph{Data availability}
The \collide{} dataset~\cite{Moreno:2025collide} is publicly available.

\begin{appendix}
\numberwithin{equation}{section}

\section{Process lists and class split}
\label{app:classes}
Tables~\ref{tab:classes-sm} and~\ref{tab:classes-bsm} list the SM and BSM processes with the event counts of the simulated samples before the selection of Sec.~\ref{sec:preproc}.

\input{tables_v3_single/process_table.tex}

\FloatBarrier

\section{Training details}
\label{app:training}

\paragraph{Preprocessing.}
The features $\pt$, $d_0$ and $d_z$ are transformed with a signed $\log(1+|x|)$. The continuous features are then standardised with the mean and standard deviation of the training sample of each view, and the type flags are left unscaled. After the selection, events have on average 3.9 L1 candidates for \minbias{}, 19 for BSM and 26 for SM (SM median 23). Because 30\% of SM and 15\% of BSM events exceed 32 candidates, the student reads on average 21 (SM) and 16 (BSM).

\paragraph{Evaluation in the appendices.}
Unless a caption says otherwise, the comparisons in Appendices~\ref{app:model-details} and~\ref{app:diagnostics} use the 10\,000-event \minbias{} sample of Sec.~\ref{sec:split}, split 21 times into 4\,000 reference, 3\,000 calibration and 3\,000 test events. The two-stage tables quote \minbias{} rates on these test events.

\paragraph{Teacher (Stage A).}
Early stopping (enabled after epoch 20, patience 15 epochs) never triggered, so we use the model after the last epoch. The validation loss falls from 0.33 to 0.23 over training.

\paragraph{Students (Stage B).}
We train the student of Table~\ref{tab:arch} as described in Secs.~\ref{sec:stageB} and~\ref{sec:hgq}, with 33\,006 steps per epoch on one NVIDIA H200 GPU. An epoch takes 230 to 400\,s. The HGQ quantisers start at 8 bits, with 1 integer bit for weights and 6 fractional bits for activations, and a monotone L1 regulariser of strength $10^{-8}$ acts on the bit widths. Weights wrap around on overflow, and values inside the attention blocks saturate with rounding. The $\beta$ controller acts on the logarithmic EBOP error with proportional gain 4 and integral gain 0.1. The EBOP estimate starts near $7\times10^{6}$ and reaches the target after 183 epochs. The selected checkpoint is epoch 216, with $4.93\times10^{5}$ EBOPs.

\paragraph{Attention quantisers.}
Appendix~\ref{app:teacher-ablation} compares two configurations of the quantisers inside the attention blocks. In KBI the data-path quantisers are parameterised by sign, total bits and integer bits with a floor of one bit, and the softmax look-up tables are fixed at five bits with two integer bits. KIF follows Ref.~\cite{Laatu:2025tft}. Its data-path quantisers are parameterised by sign, integer bits and fractional bits, with floors of zero integer and one fractional bit, and its softmax precisions are learned. The headline student uses KBI.

\paragraph{Ablation runs.}
Each teacher configuration is trained once on the same balanced subset. Small teachers have four heads and feed-forward width 16, and early stopping ends their training after 40 epochs. Students reading 64 or 128 particles use the same EBOP target, with a higher upper bound on $\beta$ for 128 particles.

\section{Ablation studies}
\label{app:model-details}
\subsection{What the teacher contributes}
\label{sec:res-ablation}
Table~\ref{tab:teacher} tests the benefit of offline-view distillation with the protocol of Sec.~\ref{sec:res-minbias}. It compares the headline student with the same student distilled from a large L1-view teacher and with the same student trained on the seen-class labels without a teacher. Without a teacher, the student retains less of the seen SM and BSM processes at the \minbias{} gate. Its classes are less separable, more SM events pass the novelty gate of Sec.~\ref{sec:res-discovery}, and the long-lived and dilepton populations do not form separate clusters. The L1-distilled student retains slightly more seen SM events and has the most separable classes. It passes the fewest \minbias{} events through both gates. However, it retains fewer BSM events at the gate, keeps fewer long-lived and dimuon events after both gates and forms no clusters. Only the student distilled from the FullReco teacher combines a high gate retention, low SM leakage through both gates and distinct clusters for both BSM populations. The student trained without a teacher settles at $2.3\times10^{5}$ EBOPs, below the target of the others.

\input{tables_v3_single/teacher_contribution_table.tex}

\subsection{Teacher view and capacity}
\label{app:teacher-ablation}
We test whether distillation from the offline view gives a better L1 embedding than training on the L1 view alone. We vary the teacher's view, FullReco or L1, and its capacity. The large teacher has width 128 and six blocks (Sec.~\ref{sec:arch}), and the small teacher has the width and depth of the student, 16 and two blocks, in floating point. The student architecture, training sample and resource budget stay fixed. An L1-trained teacher reads the 128 leading L1 particles. We train each teacher once and two students from it, one with the headline attention quantisers (KBI) and one with the alternative (KIF, Appendix~\ref{app:training}). Table~\ref{tab:ablation} gives the results on the 10\,000-event \minbias{} sample.

\input{tables_v3_single/ablation_table.tex}

With large teachers, the FullReco and L1 views give seen-SM retentions within three points of each other. The FullReco teacher is better on unseen SM and BSM processes with both quantiser configurations, so the offline view helps where generalisation is tested. Each configuration is represented by one teacher and one student, so differences of about a point should not be over-interpreted. With a small teacher the ordering reverses, by up to 7 points on SM and 3 to 9 points on BSM processes, plausibly because a small FullReco teacher sets a poor target built from information absent at L1. We have not tested this explanation directly.

Students reading 64 or 128 instead of 32 particles, trained with the same teachers and EBOP target, change the median retention over all SM processes by $-4$ to $+3$ points without a consistent gain, and not every 128-particle student reaches the target within its training. We therefore use the 32-particle student, the cheapest to implement, throughout.

\FloatBarrier

\subsection{A second split}
\label{sec:res-split48}
The whole chain was repeated with a second random split of the SM processes, again with \minbias{} unseen and with its own teacher and student. Fourteen of its 25 seen classes are also seen in the main split. Table~\ref{tab:split48} compares the two splits on the same evaluation events.

At the \minbias{} gate the second split reproduces the main split. The median retention is 81 to 96\% in every group at 1\% acceptance. The seen and unseen groups swap order, because four of the five SM processes with the lowest retention are unseen in this split ($\Upsilon\to\ell\ell$, the two Drell--Yan samples and tri-photon production). $H_{\mathrm{T}}$, which involves no training, shows the same reversal, so the ordering comes from the composition of the groups. Against the baselines of Sec.~\ref{sec:baselines} the picture of Sec.~\ref{sec:res-minbias} is unchanged. The embedding retains more than $H_{\mathrm{T}}$, the autoencoder, the input-level score and the multiplicity in every group. The hand-built feature space comes closest on BSM processes (79\% against 81\% for the embedding) and is lower on SM. In the labelled-reference probe of Sec.~\ref{sec:res-structure} the unseen classes remain locally separable.

The two-stage selection depends more on the split than the gate does. It passes 0.7\% of \minbias{} events and up to 11\% of one SM process, VBF $H\to\gamma\gamma$ (Table~\ref{tab:split48}). It keeps more of the long-lived samples than in the main split but removes most of the prompt high-mass dimuon events, and clusters of 82\% long-lived and 95\% prompt-dilepton purity still form. Which BSM signatures look unlike every seen process therefore depends on which SM processes are seen.

\input{tables_v3_single/split48_table.tex}

\FloatBarrier
\section{Additional diagnostics}
\label{app:diagnostics}

\paragraph{Baseline implementations.}
The 18 hand-built features are the logarithms of $H_{\mathrm{T}}$, multiplicity, the four leading $\pt$ values and missing $H_{\mathrm{T}}$. They also include the multiplicity and scalar $\pt$ sum for each of the five particle types and the largest transverse impact parameter of a charged hadron. Each feature, or each coordinate of the particle-input baseline, is standardised on the \minbias{} reference set. The particle-input autoencoder has 265\,296 floating-point parameters and layer widths
\[448\to256\to64\to16\to64\to256\to448.\]
It is trained with early stopping on 297\,722 \minbias{} events from files not used for its calibration or evaluation, using the student input normalisation and mean squared reconstruction error. It is an unconstrained dense autoencoder inspired by Refs.~\cite{Govorkova:2021utb,Gandrakota:2024axo} and does not reproduce their trigger implementations.

\subsection{Variations of the \minbias{} gate}
On the 10\,000-event \minbias{} sample, repeating the gate with $k=5$, 10, 20 and 50 neighbours changes the median retentions at 1\% acceptance by at most 1.1 points, with a measured test acceptance of 0.97 to 1.00\%.

The embedding score can be combined with $H_{\mathrm{T}}$ by taking, for each event, the smaller of the two calibration $p$-values and calibrating the result like any other score. The combination retains 89\%, 97\% and 86\% of the seen SM, unseen SM and BSM processes at 1\% acceptance (measured acceptance 1.01\%), within one point of the embedding alone.

An autoencoder with layers $18\to32\to8\to32\to18$ on the 18 hand-built event features of Sec.~\ref{sec:baselines}, trained on the same \minbias{} events with early stopping, retains 50\%, 70\% and 49\% of seen SM, unseen SM and BSM processes at 1\% acceptance. The nearest-neighbour distance in the same feature space retains more (76\%, 89\% and 79\%), so the reconstruction error is the weaker outlier score on event-level features, as it is on the particle-level inputs of Sec.~\ref{sec:res-minbias}.

\subsection{Local and global structure of the embedding}
\label{app:global}
\begin{figure}[tbp]
  \centering
  \includegraphics[width=\linewidth]{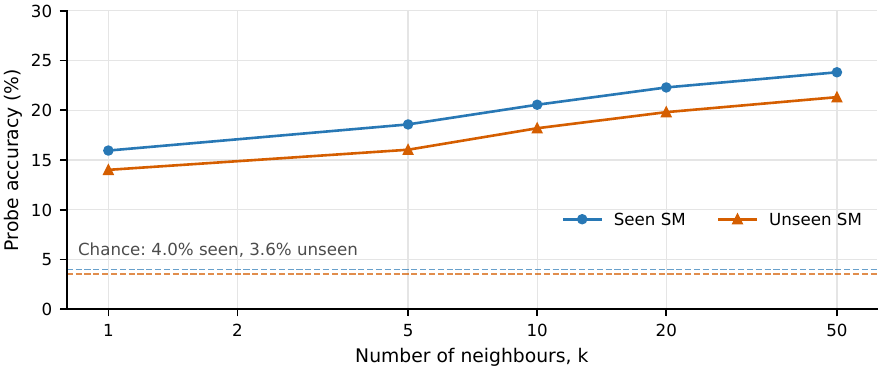}
  \caption{Labelled-reference nearest-neighbour accuracy in the embedding, evaluated separately within the seen and within the unseen group as a function of the number of neighbours $k$. Points show the mean accuracy over 21 splits of 500 reference and 500 query events per class, and dashed lines show chance.}
  \label{fig:separability}
\end{figure}

The labelled-reference probe of Sec.~\ref{sec:res-structure} shows that the classes are locally but not globally separated (Fig.~\ref{fig:separability}). Their silhouette scores~\cite{Rousseeuw:1987sil} are slightly negative, $-0.05$ for both seen and unseen classes, so an event is typically as close to the nearest other class as to its own. The two leading principal components of the seen events carry 28\% of the variance. Label-blind HDBSCAN on 200 events per process finds three regions, assigns 87\% of events to noise and has an adjusted mutual information of 0.06 with the true labels. Physically similar processes therefore overlap in the embedding and do not form separate clusters.

\subsection{Event orientation in the embedding}
\label{app:orientation}
The embedding records which way an event points along the beam axis. A logistic regression on the student's 128-dimensional output recovers the sign of the $\pt$-weighted mean $\eta$ of the event's L1 candidates for 91.3\% of events (chance 50\%), and a regression on the teacher's embedding of the same events recovers it for 81.3\%. Every process is symmetric under this reflection. The orientation is nevertheless useful in Stage~B, because it rules out about half of the 128 candidates in a batch. Mirroring an event (reversing $\eta$ and $d_z$ of every candidate) changes its embedding, with a median cosine similarity of 0.40 between an event and its mirror image.

\begin{figure}[tbp]
  \centering
  \includegraphics[width=\linewidth]{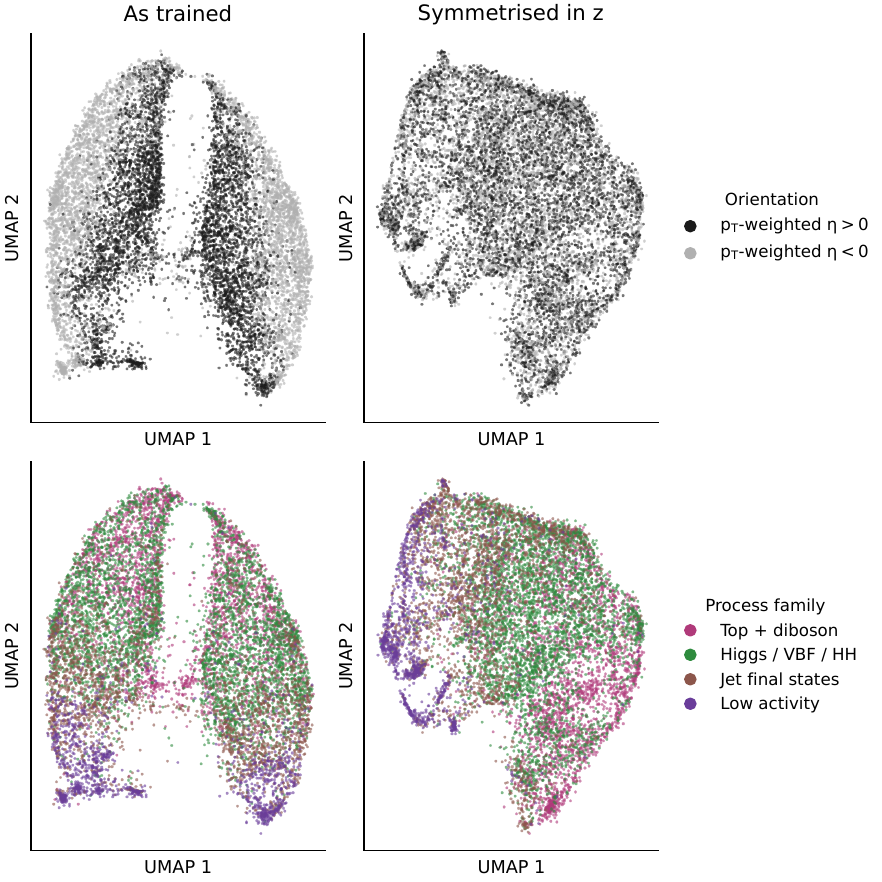}
  \caption{UMAP~\cite{McInnes:2018umap} projections (30 neighbours, minimum distance 0.1, fitted on seen events) of the student embedding as trained (left) and with each event averaged with its mirror image along the beam axis (right), for the 200 events per SM process of Fig.~\ref{fig:umap}. In the top row events are coloured by their orientation, the sign of the $\pt$-weighted mean $\eta$ of their L1 candidates, and in the bottom row by process family. As trained, the map splits by orientation into two mirror-image lobes, each containing every family. After symmetrising, the orientations mix and each family forms one region.}
  \label{fig:orientation}
\end{figure}

Averaging the normalised embeddings of each event and of its mirror image removes the orientation, and each process family then forms one region of the map (Fig.~\ref{fig:orientation}, Table~\ref{tab:orientation}). Figure~\ref{fig:umap} uses this symmetrised embedding. All trigger results use the embedding as trained, which needs only one encoder evaluation per event.

\input{tables_v3_single/orientation_table.tex}

Mirroring each view of a training pair independently with probability one half removes the incentive to encode orientation, at no cost at inference. A student trained this way, with the same teacher, data and seed, no longer encodes the orientation, and its embeddings of an event and of its mirror image coincide. It matches the headline student on SM processes at the \minbias{} gate, but retains fewer BSM events there and keeps far fewer prompt high-mass dimuon events after both gates (Table~\ref{tab:orientation}). The headline results therefore use the student without augmentation.

\subsection{Scoring and sensitivity of the clustering}
\label{app:discovery}
Labels are used only to score the clustering. For each true process $c$, let $n_c$ be the number of its events retained by both gates. For each cluster $C$, the purity is $P=|c\cap C|/|C|$ and the recovery is $R=|c\cap C|/n_c$. The process is credited with the cluster that maximises
\begin{equation}
  F_1 = \frac{2PR}{P+R}.
  \label{eq:f1}
\end{equation}
A process is called a \emph{discovery candidate} if it retains at least 10\% of its events after both gates and its best cluster has $F_1\ge0.3$. This descriptive criterion does not measure statistical significance.
Two prompt high-mass dimuon samples pass the candidate criterion, with $F_1$ of 0.33 and 0.42 and recoveries of at least 96\% (Table~\ref{tab:discovery-candidates}). The purity and $F_1$ of each sample are modest because several samples share each signature.

We repeated the clustering step with every combination of 16, 32 and 64 principal components, minimum cluster sizes of 10, 20 and 40, and minimum-sample values of 5, 10 and 20. The student's 128-dimensional output lies in a 16-dimensional affine subspace, so 16 or more components keep essentially all its variance and the number of components has little effect. The two prompt dimuon candidates pass the criterion at all 27 settings, and the hidden-valley dimuon sample with $c\tau=100$\,mm passes at 23. A cluster at least 90\% prompt dilepton forms at all 27 settings, whereas a cluster at least 85\% long-lived forms at only 3. The two gates, and hence the retained fractions of Fig.~\ref{fig:discovery-repeats}, do not depend on these settings.

\input{tables_v3_single/discovery_candidates_table.tex}

\subsection{Paired teacher--student alignment}
\label{app:alignment}
The teacher and the student were evaluated on the same 1\,000 events per SM process, so their embeddings are paired event by event. The mean cosine similarity between the student and teacher embeddings of the same event is 0.439 for seen and 0.441 for unseen processes. Re-pairing each student event with a different teacher event of the same process reduces these means to 0.232 and 0.224. This control preserves process identity and removes event identity, so the drop measures the event-level information that Stage~B transfers.

We also rank each student event's paired teacher event among 128 teacher candidates from the same process, so that only event-level information can help. The paired event ranks first for 39.0\% of seen and 40.1\% of unseen events, against a random expectation of 0.8\%. Unseen processes match or exceed seen ones, so the distillation also generalises at the level of single events.

\subsection{Per-process retention}

\begin{figure}[tbp]
  \centering
  \includegraphics[width=\linewidth,height=0.9\textheight,keepaspectratio]{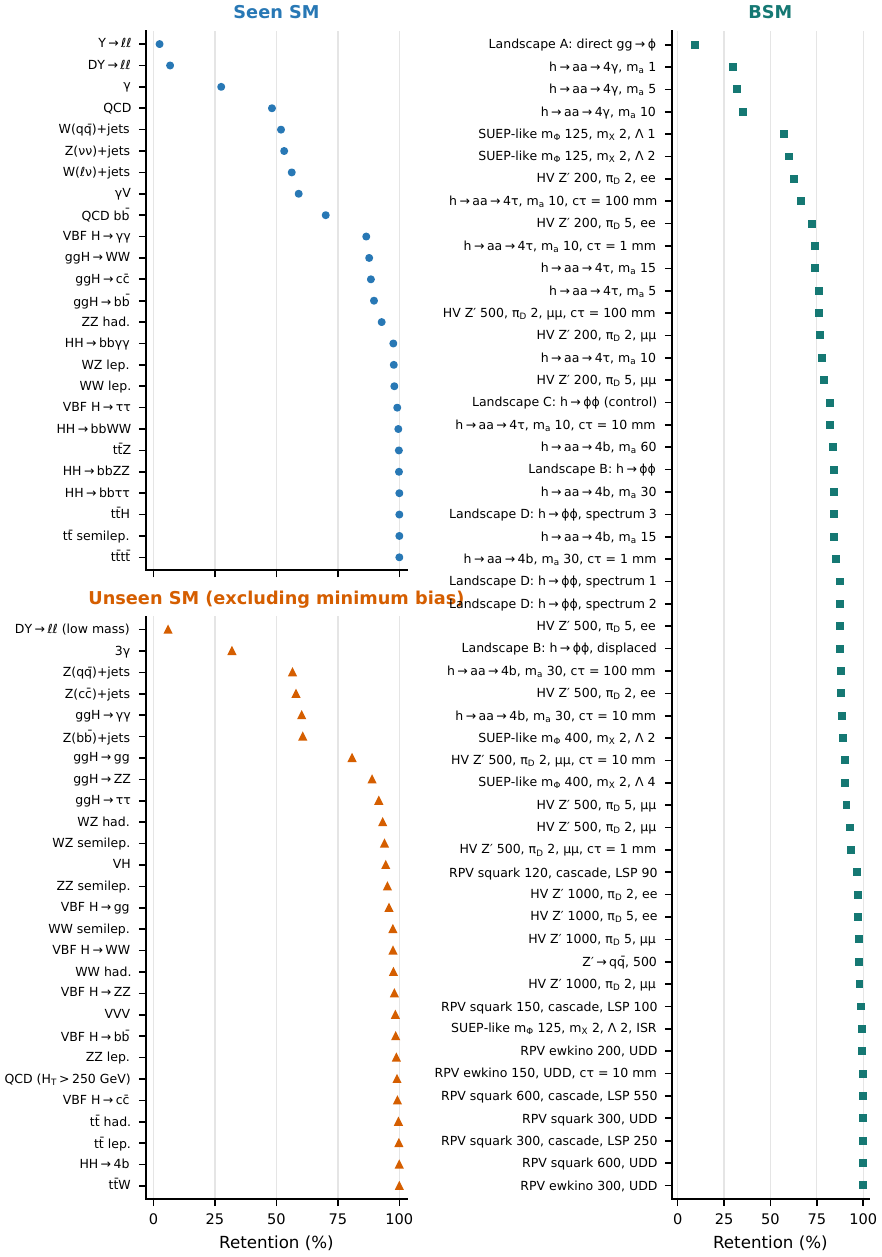}
  \caption{Per-process retention at a target \minbias{} acceptance of 1\%, averaged over the ten splits of the large \minbias{} sample. Processes are ordered by retention within each group, and generator suffixes are omitted from the names.}
  \label{fig:minbias-processes}
\end{figure}

\FloatBarrier

\end{appendix}

\bibliography{references}

\end{document}

%% file: tables_v3_single/baselines_table.tex
\begin{table}[tbp]
\centering\small
\begin{tabular}{lccc}
\toprule
Score at 1\% \minbias{} acceptance & Seen SM (\%) & Unseen SM (\%) & BSM (\%) \\
\midrule
L1 embedding (this work) & \textbf{90} & \textbf{96} & \textbf{88} \\
Same student, no teacher & 86 & 95 & 85 \\
Hand-built features, nearest neighbours & 78 & 91 & 82 \\
$H_{\mathrm{T}}$ & 86 & \textbf{96} & 75 \\
Autoencoder & 70 & 79 & 51 \\
Student inputs, nearest neighbours & 70 & 79 & 46 \\
Multiplicity & 70 & 78 & 30 \\
\bottomrule
\end{tabular}
\caption{Median process retention at 1\% \minbias{} acceptance for the L1 embedding, the same student trained without a teacher, and five trigger baselines (Sec.~\ref{sec:baselines}), all calibrated on the large \minbias{} sample.}
\label{tab:baselines}
\end{table}

%% file: tables_v3_single/resource_comparison_table.tex
\begin{table}[t]
\centering\small\setlength{\tabcolsep}{6pt}
\begin{tabular}{lrrrrrr}
\toprule
Model & \shortstack{Latency\\(ns)} & \shortstack{LUT\\(k)} & FF & DSP & BRAM & \shortstack{II\\(clk)} \\
\midrule
Linformer & 147 & 239.4 & 99\,957 & 12 & 0 & 1 \\
\bottomrule
\end{tabular}
\caption{Post-route resources of the encoder on a VU13P. LUT counts are in thousands; FF, DSP and BRAM are absolute counts.}
\label{tab:resources}
\end{table}

%% file: tables_v3_single/process_table.tex
\begin{table}[p]
\centering\small
\begin{tabular}{lr@{\hspace{1.5em}}lr}
\toprule
\multicolumn{2}{l}{Seen SM (25)} & \multicolumn{2}{l}{Unseen SM (28)} \\
Process & Events & Process & Events \\
\midrule
\texttt{DYJetsToLL} & 499\,661  & \texttt{HH\_4b} & 1\,499\,060  \\
\texttt{HH\_bbWW} & 1\,220\,458  & \texttt{VBFHWW} & 1\,164\,956  \\
\texttt{HH\_bbZZ} & 1\,238\,527  & \texttt{VBFHZZ} & 1\,161\,110  \\
\texttt{HH\_bbgammagamma} & 1\,279\,246  & \texttt{VBFHbb} & 1\,265\,403  \\
\texttt{HH\_bbtautau} & 1\,289\,170  & \texttt{VBFHcc} & 1\,136\,392  \\
\texttt{QCD\_HT50toInf} & 1\,158\,265  & \texttt{VBFHgluglu} & 1\,393\,108  \\
\texttt{QCD\_HT50tobb} & 1\,334\,765  & \texttt{VH\_incl} & 1\,491\,781  \\
\texttt{VBFHgammagamma} & 1\,156\,924  & \texttt{VVV\_incl} & 1\,220\,691  \\
\texttt{VBFHtautau} & 1\,135\,487  & \texttt{WW\_hadronic} & 1\,394\,918  \\
\texttt{WJetsToLNu} & 1\,453\,749  & \texttt{WW\_semileptonic} & 1\,404\,606  \\
\texttt{WJetsToQQ} & 1\,477\,306  & \texttt{WZ\_hadronic} & 1\,329\,343  \\
\texttt{WW\_leptonic} & 1\,441\,092  & \texttt{WZ\_semileptonic} & 1\,352\,558  \\
\texttt{WZ\_leptonic} & 1\,411\,754  & \texttt{ZJetsToQQ} & 1\,486\,935  \\
\texttt{ZJetsTovv} & 1\,481\,134  & \texttt{ZJetsTobb} & 1\,486\,973  \\
\texttt{ZZ\_hadronic} & 1\,324\,962  & \texttt{ZJetsTocc} & 1\,496\,896  \\
\texttt{gamma} & 1\,318\,162  & \texttt{ZZ\_leptonic} & 1\,380\,161  \\
\texttt{gamma\_V} & 1\,432\,252  & \texttt{ZZ\_semileptonic} & 1\,344\,927  \\
\texttt{ggHWW} & 1\,498\,590  & \texttt{ggHZZ} & 1\,498\,562  \\
\texttt{ggHbb} & 1\,499\,029  & \texttt{ggHgammagamma} & 1\,428\,695  \\
\texttt{ggHcc} & 1\,458\,757  & \texttt{ggHgluglu} & 1\,499\,076  \\
\texttt{ttH\_incl} & 1\,410\,706  & \texttt{ggHtautau} & 1\,488\,506  \\
\texttt{ttZ\_incl} & 1\,431\,410  & \texttt{minbias} & 1\,500\,000  \\
\texttt{tt\_semiLeptonic} & 1\,462\,009  & \texttt{tri\_gamma} & 199\,981  \\
\texttt{tttt\_incl} & 1\,470\,000  & \texttt{ttW\_incl} & 1\,416\,569  \\
\texttt{upsilon\_to\_leptons} & 1\,500\,000  & \texttt{tt\_hadronic} & 1\,464\,185  \\
&  & \texttt{tt\_leptonic} & 1\,462\,054  \\
&  & \texttt{DYJetsToLL\_lowmass}$^{*}$ & 999\,365  \\
&  & \texttt{QCD\_HT250toInf}$^{*}$ & 880\,072  \\
\bottomrule
\end{tabular}
\caption{The 53 SM processes and their random split into seen and unseen processes (Sec.~\ref{sec:split}). Event counts are those of the samples used here, before selection, and generator suffixes are omitted from the names. $^{*}$Added after training and unseen for every model.}
\label{tab:classes-sm}
\end{table}

\begin{table}[p]
\centering\footnotesize
\begin{tabular}{ll}
\toprule
\multicolumn{2}{l}{BSM benchmarks (52; about 50\,000 events each; evaluation only)} \\
\midrule
\texttt{HVdilep\_Zp1000\_piD2\_ee} & \texttt{SUEPlike\_HV\_mPhi125\_mX2\_Lam2\_ISR} \\
\texttt{HVdilep\_Zp1000\_piD2\_mumu} & \texttt{SUEPlike\_HV\_mPhi400\_mX2\_Lam2} \\
\texttt{HVdilep\_Zp1000\_piD5\_ee} & \texttt{SUEPlike\_HV\_mPhi400\_mX2\_Lam4} \\
\texttt{HVdilep\_Zp1000\_piD5\_mumu} & \texttt{Zprime\_qq\_m500} \\
\texttt{HVdilep\_Zp200\_piD2\_ee} & \texttt{hToAA\_4b\_ma15} \\
\texttt{HVdilep\_Zp200\_piD2\_mumu} & \texttt{hToAA\_4b\_ma30} \\
\texttt{HVdilep\_Zp200\_piD5\_ee} & \texttt{hToAA\_4b\_ma30\_ctau100mm} \\
\texttt{HVdilep\_Zp200\_piD5\_mumu} & \texttt{hToAA\_4b\_ma30\_ctau10mm} \\
\texttt{HVdilep\_Zp500\_piD2\_ee} & \texttt{hToAA\_4b\_ma30\_ctau1mm} \\
\texttt{HVdilep\_Zp500\_piD2\_mumu} & \texttt{hToAA\_4b\_ma60} \\
\texttt{HVdilep\_Zp500\_piD2\_mumu\_ctau100mm} & \texttt{hToAA\_4gamma\_ma1} \\
\texttt{HVdilep\_Zp500\_piD2\_mumu\_ctau10mm} & \texttt{hToAA\_4gamma\_ma10} \\
\texttt{HVdilep\_Zp500\_piD2\_mumu\_ctau1mm} & \texttt{hToAA\_4gamma\_ma5} \\
\texttt{HVdilep\_Zp500\_piD5\_ee} & \texttt{hToAA\_4tau\_ma10} \\
\texttt{HVdilep\_Zp500\_piD5\_mumu} & \texttt{hToAA\_4tau\_ma10\_ctau100mm} \\
\texttt{RPV\_ewkino150\_UDD\_ctau10mm} & \texttt{hToAA\_4tau\_ma10\_ctau10mm} \\
\texttt{RPV\_ewkino200\_UDD} & \texttt{hToAA\_4tau\_ma10\_ctau1mm} \\
\texttt{RPV\_ewkino300\_UDD} & \texttt{hToAA\_4tau\_ma15} \\
\texttt{RPV\_squark120\_cascade\_LSP90} & \texttt{hToAA\_4tau\_ma5} \\
\texttt{RPV\_squark150\_cascade\_LSP100} & \texttt{landscape\_LA\_direct} \\
\texttt{RPV\_squark300\_UDD} & \texttt{landscape\_LB\_higgs} \\
\texttt{RPV\_squark300\_cascade\_LSP250} & \texttt{landscape\_LB\_higgs\_displaced} \\
\texttt{RPV\_squark600\_UDD} & \texttt{landscape\_LC\_control} (control) \\
\texttt{RPV\_squark600\_cascade\_LSP550} & \texttt{landscape\_LD\_spectrum1} \\
\texttt{SUEPlike\_HV\_mPhi125\_mX2\_Lam1} & \texttt{landscape\_LD\_spectrum2} \\
\texttt{SUEPlike\_HV\_mPhi125\_mX2\_Lam2} & \texttt{landscape\_LD\_spectrum3} \\
\bottomrule
\end{tabular}
\caption{The 52 BSM benchmark samples. \texttt{HVdilep} samples are hidden-valley $Z^{\prime}\to$ dark-pion $\to$ dilepton models labelled by the $Z^{\prime}$ mass in GeV, the dark-pion mass in GeV and the lepton flavour. \texttt{RPV} samples are $R$-parity-violating supersymmetry models with UDD couplings or cascade decays, labelled by the squark or electroweakino mass in GeV. \texttt{SUEPlike} samples model soft unclustered energy patterns from a dark shower and are labelled by the parameters $m_\Phi$, $m_X$ and $\Lambda$. \texttt{hToAA} samples are exotic Higgs decays to pairs of light pseudoscalars of the stated mass in GeV, and \texttt{Zprime\_qq} is a 500\,GeV $Z^{\prime}\to q\bar{q}$ resonance. The suffix \texttt{ctau} gives the proper decay length of the displaced samples. The \texttt{landscape} samples are light-scalar models of Ref.~\cite{DAgnolo:2025cxb}, with direct gluon-fusion production (LA), $h\to\phi\phi$ with prompt or displaced decays (LB), the same with three other mass spectra (LD), and a control sample with the hidden channel closed (LC). Figures abbreviate the names, for example HV $Z^{\prime}$ 500, $\pi_D$ 2, $\mu\mu$, $c\tau=100$\,mm for \texttt{HVdilep\_Zp500\_piD2\_mumu\_ctau100mm}.}
\label{tab:classes-bsm}
\end{table}

%% file: tables_v3_single/teacher_contribution_table.tex
\begin{table}[tbp]
\centering\small\setlength{\tabcolsep}{5pt}
\begin{tabular}{lccc}
\toprule
 & FullReco-distilled & L1-distilled & No teacher \\
\midrule
\multicolumn{4}{l}{\emph{Median retention at 1\% \minbias{} acceptance (\%)}} \\
\quad Seen SM & 90 & \textbf{91} & 86 \\
\quad Unseen SM & \textbf{96} & 95 & 95 \\
\quad BSM & \textbf{88} & 85 & 85 \\
\midrule
\multicolumn{4}{l}{\emph{Two-stage selection (\%)}} \\
\quad \minbias{} passing both gates & 0.33 & \textbf{0.00} & 0.43 \\
\quad Highest SM retention & \textbf{3.3} & 4.0 & 6.3 \\
\quad Long-lived samples & \textbf{23.8} & 8.2 & 13.1 \\
\quad High-mass dimuon samples & \textbf{25.8} & 0.3 & 2.0 \\
\quad Purest long-lived cluster & \textbf{80} & -- & 23 \\
\quad Purest prompt-dilepton cluster & \textbf{97} & -- & 15 \\
\bottomrule
\end{tabular}
\caption{The same 32-particle student distilled from a six-block, width-128 teacher using the FullReco or L1 view, and trained without a teacher, with the large \minbias{} sample of Sec.~\ref{sec:trigger}. The two-stage rows are defined as in Table~\ref{tab:orientation}, and a dash means that no cluster forms.}
\label{tab:teacher}
\end{table}

%% file: tables_v3_single/ablation_table.tex
\begin{table}[t]
\centering\small
\begin{tabular}{llcccc}
\toprule
Teacher view & Teacher size & Seen SM & Unseen SM & BSM & All SM \\
\midrule
\multicolumn{6}{l}{\emph{KBI attention quantisers}} \\
FullReco & large & 89 & \textbf{96} & \textbf{86} & \textbf{94} \\
L1 & large & \textbf{90} & 95 & 82 & 93 \\
FullReco & small & 76 & 90 & 64 & 86 \\
L1 & small & 83 & 93 & 73 & 90 \\
\midrule
\multicolumn{6}{l}{\emph{KIF attention quantisers}} \\
FullReco & large & \textbf{90} & \textbf{96} & \textbf{86} & \textbf{95} \\
L1 & large & 88 & 92 & 79 & 91 \\
FullReco & small & 78 & 91 & 70 & 86 \\
L1 & small & 85 & 93 & 73 & 91 \\
\bottomrule
\end{tabular}
\caption{Median process retention (\%) at a 1\% calibrated \minbias{} acceptance for 32-particle students distilled from four teachers, averaged over 21 splits of the 10\,000-event \minbias{} sample. ``All SM'' pools the 52 non-\minbias{} SM processes.}
\label{tab:ablation}
\end{table}

%% file: tables_v3_single/split48_table.tex
\begin{table}[tbp]
\centering\small
\begin{tabular}{lcc}
\toprule
& Main split & Second split \\
\midrule
\multicolumn{3}{l}{\emph{Median retention at 1\% \minbias{} acceptance (\%)}} \\
\quad Seen SM & 89 & 96 \\
\quad Unseen SM & 96 & 92 \\
\quad BSM & 86 & 81 \\
\quad Measured \minbias{} acceptance & 0.99 & 1.00 \\
\midrule
\multicolumn{3}{l}{\emph{Two-stage selection (\%)}} \\
\quad \minbias{} passing both gates & 0.33 & 0.73 \\
\quad Highest SM retention & 3.3 & 11.0 \\
\quad Long-lived samples & 23.8 & 43.5 \\
\quad High-mass dimuon samples & 25.8 & 2.5 \\
\quad Purest long-lived cluster & 80 & 82 \\
\quad Purest prompt-dilepton cluster & 97 & 95 \\
\bottomrule
\end{tabular}
\caption{The main and second splits into seen and unseen processes, evaluated on the same events and the 10\,000-event \minbias{} sample. The two-stage rows are defined as in Table~\ref{tab:orientation}.}
\label{tab:split48}
\end{table}

%% file: tables_v3_single/orientation_table.tex
\begin{table}[tbp]
\centering\small
\begin{tabular}{lccc}
\toprule
& As trained & Averaged after & Mirrored in \\
& & training & training \\
\midrule
Orientation decodable (\%, chance 50\%) & 91.3 & \textbf{50.5} & 52.2\\
\midrule
\multicolumn{4}{l}{\emph{Probe accuracy (\%)}} \\
\quad Seen SM & 20.6 & \textbf{21.7} & 21.1\\
\quad Unseen SM & 18.2 & \textbf{19.8} & 18.7\\
\midrule
\multicolumn{4}{l}{\emph{Median retention at 1\% \minbias{} acceptance (\%)}} \\
\quad Seen SM & 89 & \textbf{91} & 90\\
\quad Unseen SM & \textbf{96} & \textbf{96} & \textbf{96}\\
\quad BSM & 86 & \textbf{87} & 84\\
\midrule
\multicolumn{4}{l}{\emph{Two-stage selection (\%)}} \\
\quad \minbias{} passing both gates & 0.33 & 0.30 & \textbf{0.10}\\
\quad Highest SM retention & \textbf{3.3} & 5.0 & 7.0\\
\quad Long-lived samples & 23.8 & \textbf{25.0} & 14.0\\
\quad High-mass dimuon samples & \textbf{25.8} & 19.2 & 1.8\\
\quad Purest long-lived cluster & \textbf{80} & 65 & 74\\
\quad Purest prompt-dilepton cluster & \textbf{97} & 96 & 96\\
\bottomrule
\end{tabular}
\caption{Event orientation and its removal for the headline student as trained, with each event averaged with its mirror image, and for a student trained with randomly mirrored views (same teacher, data and seed), on the 10\,000-event \minbias{} sample. Long-lived samples have $c\tau=10$ or 100\,mm. High-mass dimuon samples are the two prompt hidden-valley $Z^{\prime}\to\mu\mu$ samples at 1\,TeV. The cluster rows give the largest fraction of long-lived or of prompt hidden-valley dilepton events in any HDBSCAN cluster at the default settings. Highest SM retention is the largest fraction of any SM process passing both gates.}
\label{tab:orientation}
\end{table}

%% file: tables_v3_single/discovery_candidates_table.tex
\begin{table}[tbp]
\centering\footnotesize\setlength{\tabcolsep}{4pt}
\begin{tabular}{rlrrrrr}
\toprule
No. & BSM benchmark & Retained (\%) & Cluster size & Purity & Recovery & $F_1$ \\
\midrule
1 & \texttt{HVdilep\_Zp1000\_piD5\_mumu} & 29 & 327 & 0.26 & 0.99 & 0.42 \\
2 & \texttt{HVdilep\_Zp1000\_piD2\_mumu} & 23 & 327 & 0.20 & 0.96 & 0.33 \\
\bottomrule
\end{tabular}
\caption{Discovery candidates, the processes with at least 10\% of their 300 stream events retained after both gates and a best-matching cluster with $F_1\ge0.3$. Purity, recovery and $F_1$ are defined in Appendix~\ref{app:discovery}.}
\label{tab:discovery-candidates}
\end{table}